\documentclass{PoS}

\usepackage[T1]{fontenc}
\usepackage{bbding} 
\usepackage{amsmath}
\usepackage{graphicx,float,color}
\usepackage{braket}
\usepackage{amsfonts}
\usepackage{amssymb} 
\usepackage{epic}
\usepackage{eepic}
\usepackage{epsfig}
\usepackage{latexsym}
\usepackage{caption}
\usepackage[caption=false]{subfig}
\usepackage{float}
\usepackage{bm}
\usepackage{slashed}

\def\Mp{m_{\mathrm{Pl}}}
\def\lp{\ell_{\mathrm{Pl}}}

\def\Mmin{M_{\mathrm{min}}}
\def\hMmin{\hat{M}_{\mathrm{min}}}
\def\Mini{M_{\mathrm{init}}}
\def\hMini{\hat{M}_{\mathrm{init}}}
\def\Mpeak{M_\textrm{peak}}

\title{\center Primordial Black-Hole Mimicker in Quadratic Gravity \\as Dark Matter}

\ShortTitle{Primordial Black-Hole Mimicker in Quadratic Gravity as Dark Matter}

\author{\speaker{Ufuk Aydemir}
\thanks{This article is the write-up of an extended version of the talk given at CORFU2019 and reports mainly on the work done in collaboration with Bob Holdom and Jing Ren~\cite{Aydemir:2020xfd}. In addition; some complementary details are provided regarding the previous works on 2-2-holes, a brief comparison between black holes and black-hole mimickers is included in Introduction, and some remarks on the entropy-area law are made as a part of section II, where 2-2-holes are reviewed.} 
\\     
      Institute of High Energy Physics, Chinese Academy of Sciences, Beijing 100049, P. R. China\\   
       E-mail: \email{uaydemir@ihep.ac.cn}}

\abstract{
We discuss the astrophysical and cosmological implications of having primordial thermal 2-2-hole remnants as dark matter. Thermal 2-2-holes emanate in quadratic gravity as horizonless classical
solutions for ultracompact distributions of relativistic thermal gas. 
In contrast to a large 2-2-hole that imitates the thermodynamic behaviour of a black hole, 
a small 2-2-hole at late stages of evaporation behaves as a stable remnant with the mass approaching a minimal value. These remnants as all dark matter can satisfy the corresponding observational constraints provided that both the formation and remnant masses are relatively small.  The parameter space for the remnant mass is probed through possible remnant mergers that would produce strong fluxes of high-energy astrophysical particles; the high-energy photon and neutrino data appear to favor towards the Planck-mass remnants, pointing to the strong-coupling scenario for the quantum theory of quadratic gravity. The formation mass, on the other hand, is constrained by the early-universe cosmology, which turns out to require 2-2-holes to evolve into the remnant state before Big Bang Nucleosynthesis.
}

\FullConference{Corfu Summer Institute 2019 "School and Workshops on Elementary Particle Physics and Gravity" (CORFU2019)\\
		31 August - 25 September 2019\\
		Corfu, Greece}

\begin{document}

\section{Introduction}

\subsection{Overview}

General Relativity (GR) is an outstandingly successful theory of gravity at large scales, which provides a compelling cosmological framework, explains astrophysical phenomena successfully, and has passed every test it has so far been confronted~\cite{Will:2014kxa}. The gravitational waves, one of the main predictions of GR, have also been observed first time several years ago~\cite{Abbott:2016blz,Abbott:2016nmj}, with many events detected since then~\cite{LIGOScientific:2019fpa}. However, there is a common theoretical consensus that GR is not the end of story for gravity. A major driving force behind the search for a theory beyond GR is the determination towards a renormalizable quantum field theory as the theory of quantum gravity.
 


Another motivation to search for a theory of quantum gravity is the resolution it is anticipated to provide on the information-loss problem in black holes, likely through modifications regarding horizon formation. In general, from \textit{a priori} dimensional arguments, direct effects of quantum gravity are not expected to manifest around a macroscopic horizon since these effects are  typically associated with the Planck-size curvatures. Yet this is a theoretically viable possibility; there exist arguments from the standpoint of effective field theory~\cite{Almheiri:2012rt,Giddings:2014ova}\footnote{See also Ref.~\cite{Giddings:2012bm}.} and string theory~\cite{Mathur:2008nj} pointing to existence of near-horizon modifications as a resolution to the information-loss problem. A well-studied example is fuzzballs~\cite{Mathur:2005zp,Mathur:2008nj}, where the horizon is replaced by a stringy interface. There exist many other proposed objects without event horizons, which are commonly referred to as horizonless ultracompact objects, \textit{aka} black-hole mimickers~\cite{Cardoso:2019rvt}. Many of these objects, besides providing a resolution for the information-loss issue, constitute viable dark matter candidates. 


In this article, we focus on such an object, referred to as 2-2-hole, that arises in quadratic gravity, a  renormalizable and asymptotically free candidate for quantum gravity in the framework of quantum field theories~\cite{Stelle:1976gc, Voronov:1984kq, Fradkin:1981iu, Avramidi:1985ki}, whose action includes on top of the Einstein-Hilbert action all the quadratic curvature terms, i.e. the Weyl ($C^{\mu\nu\rho\sigma}C_{\mu\nu\rho\sigma}$) and Ricci ($R^2$) terms.\footnote{It has been long-known that the theory at the classical level suffers from a ghost problem associated with the higher derivative terms. There is a long list of proposed solutions to deal with the ghost by taking quantum corrections seriously~\cite{Lee:1969fy, Tomboulis:1977jk, Grinstein:2008bg, Anselmi:2017yux, Donoghue:2018lmc, Bender:2007wu, Salvio:2015gsi, Holdom:2015kbf, Holdom:2016xfn, Salvio:2018crh}. In this work, we remain agnostic on the possible resolution of this issue.}  A 2-2-hole comes out as a classical
solution in the theory when sourced by a ultracompact matter distribution~\cite{Holdom:2002xy, Holdom:2016nek,Holdom:2019ouz,Ren:2019afg}. The object  resembles a black hole in the exterior, whereas in the interior it is characterised by a distinct high-curvature solution, with a transition region at around the would-be horizon. A 2-2-hole can be arbitrarily heavy but, unlike many other ultracompact objects, has a minimum allowed mass $\Mmin$, indicating the existence of stable remnants. The remnant mass is determined by the the mass of the spin-2 mode in the theory, $\Mmin\sim \Mp^2/m_2$, hence carries information on the underlying theory of quantum gravity.  Remnants with around the Planck mass, $\Mmin\sim \Mp$, would correspond to the strong coupling scenario, whereas them being heavier would point to the weak coupling case. 

The case of a relativistic thermal gas as the matter source for 2-2-holes was investigated in \cite{Holdom:2019ouz} and  \cite{ Ren:2019afg}. It was found that the thermodynamic properties of a thermal 2-2-hole in the large-mass stage have the same form as a black hole, thus the evaporation process in this early-stage shares most of the features of the black hole evaporation; a large 2-2-hole radiates with a Hawking-like temperature and exhibits an entropy-area law. Once the temperature reaches the peak value, the 2-2-hole enters into the remnant stage with mass close to the minimal mass $\Mmin$, where the object behaves more like an ordinary thermodynamic system. In this stage, the heat capacity becomes positive, the evaporation significantly slows down and asymptotically halts. 

In this work~\cite{Aydemir:2020xfd}, 2-2-hole remnants are considered as dark matter. The non-remnant version is quite similar to the case of Primordial Black Holes (PBHs)~\cite{Carr:2009jm,Carr:2016drx,Dalianis:2019asr,Carr:2020gox} and hence it is already heavily constrained. In the standard PBH scenario, only very few narrow mass windows are still available for $M_\textrm{PBH}\gtrsim 10^{15}\,$g. Smaller PBHs are assumed to have evaporated away by the present epoch. However, it has been conjectured in the literature that the evaporation comes to a stop at some stage and a remnant is left behind that may serve as dark matter~\cite{MacGibbon:1987my,Barrow:1992hq,Carr:1994ar}. In this case, the $M_\textrm{PBH}\lesssim 10^{15}\,$g range is still allowed. In fact, it has been shown that all the observational constraints can be satisfied provided that PBHs radiate away most of their energy before Big Bang Nucleosynthesis (BBN). The Planck-mass remnants, with the initial mass satisfying $M_\textrm{PBH}\lesssim 10^{6}\,$g, can account for all of dark matter~\cite{Carr:2009jm, Dalianis:2019asr,Carr:2020gox}. 

2-2-hole remnants possess various appealing features over their black-hole counterparts. In addition to innately avoiding the information-loss problem, a stable 2-2-hole remnant arises naturally in the theory. The underlying mechanism for both stabilization and the absence of horizon is evident, stemming from quadratic curvature terms that operate at high energies or curvatures. In contrast, in the case of black-hole remnants, neither their generation mechanism nor how they resolve the information-loss problem is well-understood~\cite{Chen:2014jwq}. Moreover, black hole remnants suffer from issue of an arbitrarily large amount of  entropy stored in their relatively small-size~\cite{Giddings:1992hh}, whereas for 2-2-hole remnants there is no such problem since the information is carried out from a 2-2-hole by the thermal radiation, as with any burning object.

Furthermore,  a distinct phenomenon occurs in the case of 2-2-hole remnants enabling the small-mass range testable by direct  non-gravitational observations, unlike black-hole remnants.  A binary merger of two remnants gives rise to a high temperature, non-remnant, product with the excess energy released almost instantly by emitting high-energy particles. This strong flux can be confronted by the observations in high-energy astrophysical particles to constraint the parameter space. This is indeed what we do in this work, in addition to deriving the early-universe constraints from BBN, CMB, and dark matter relic abundance.

\subsection{Black holes vs. black-hole mimickers}
\label{sec:comparison}

Black-hole mimickers, i.e. horizonless ultracompact objects, have long been an active research area particularly in the context of information loss problem and dark matter~\cite{Cardoso:2019rvt,Raidal:2018eoo}. The detected gravitational waves appear to be consistent with stellar-mass astrophysical black holes~\cite{LIGOScientific:2019fpa}, yet implications regarding near-horizon physics are not clear, and horizonless ultracompact objects are viable alternatives to black holes. These objects in general appear just like black holes from outside, away from the (would-be) horizon, a remarkable example of which is the 2-2-hole, which converges to the Schwarzschild solution in the exterior. 2-2 holes, and many other such objects, appear dark to a distant observer, despite of the absence of horizon, due to light being trapped in the high-redshift region in deep gravitational potential~\cite{Holdom:2016nek,Cardoso:2019rvt}.

The immediate question is then whether and how these objects can be differentiated from black holes. 
Even though this appears to be a difficult task for the current observations,  there is room for exploration.
 It has been argued in the literature that horizonless ultracompact objects may leave distinctive imprints as echoes in gravitational-wave signals~\cite{Conklin:2017lwb,Conklin:2019fcs,Cardoso:2019rvt}. 
 In fact, there has been a recent discussion on the existence of such signals in the observed data~\cite{Abedi:2016hgu,Westerweck:2017hus,Abedi:2018pst}. A possible detection of such objects in future searches, and even in the observed gravitational waves by further analysing the existence data carefully, remains to be a viable possibility. With the high sensitivity of the anticipated next-generation  interferometers, the future of GW astrophysics seems promising to this end. 
  
Since horizonless ultracompact objects, in particular 2-2-holes, can shed light on crucial issues such as the information-loss  and  dark matter problems, it is an important task to understand the phenomenological implications of these objects and explore the available parameter space through observational constraints. Furthermore, any information on 2-2-holes could also provide insights on the underlying theory of quantum gravity, particularly through the constraints on $\Mmin$.
 
 
The rest of the article is organised as follows. 2-2-holes are reviewed in section~\ref{sec:general}, where we also reflect on the entropy-are law at the end. A brief introduction for primordial 2-2-holes as dark matter is given in section~\ref{sec:dmintro}. Constraints from the present-epoch observations and early universe cosmology are discussed in section~\ref{sec:pheno0}, which also includes a summary of results. The article is concluded in section~\ref{sec:final}.

\section{2-2-holes: Horizonless ultracompact structures in quadratic gravity}
\label{sec:general}
\subsection{Preliminaries: From the vacuum solutions to the thermal 2-2-holes}
In this subsection, we review 2-2-holes by following Refs.~\cite{Holdom:2002xy,Holdom:2016nek,Holdom:2019ouz,Ren:2019afg} and Ref.~\cite{Aydemir:2020xfd}. 
The action of quadratic gravity is given as 
\begin{eqnarray}
\label{action}
S_{\mathrm{QG}}= \frac{1}{16\pi}\int d^4 x \sqrt{-g}\left(m_{\mathrm{Pl}}^2 R-\alpha\; C_{\mu\nu\rho\sigma}C^{\mu\nu\rho\sigma}+\beta R^2\right)
\end{eqnarray}
where $\alpha$ and $\beta$ are dimensionless couplings, and $C_{\mu\nu\rho\sigma}$ is the Weyl tensor. In addition to the usual massless graviton, the theory describes a spin-0 and a spin-2 mode with the tree level masses  $m_0=\Mp/\sqrt{6\beta}$ and $m_2=\Mp/\sqrt{2\alpha}$, respectively. At the quantum level, the theory is renormalizable and asymptotically free~\cite{Stelle:1976gc, Voronov:1984kq, Fradkin:1981iu, Avramidi:1985ki}.  At the classical level, the theory suffers from the infamous ghost problem, associated with the spin-2 mode, due to the higher derivative terms in the action. The proposed methods to deal with this pathology mostly involve modifications to quantum prescription, depending on whether the theory becomes strongly or weakly coupled at the quantum-gravity, i.e.~the Planck, scale~\cite{Lee:1969fy, Tomboulis:1977jk, Grinstein:2008bg, Anselmi:2017yux, Donoghue:2018lmc, Bender:2007wu, Salvio:2015gsi, Holdom:2015kbf, Holdom:2016xfn, Salvio:2018crh}. There is still no consensus on the resolution of this problem.




In Ref.~\cite{Holdom:2002xy}, a novel spherically symmetric, asymptotically flat, and static vacuum solutions was investigated. These objects, recently referred to as 2-2-holes, belongs to one of the three family of solutions~\cite{Stelle:1977ry}, the others of which are black hole and star-like solutions, as we will mention in a little more detail below.

The general metric for a static, spherically symmetric spacetime is given as
\begin{eqnarray}
\label{metric}
ds^2= -B(r)\; dt^2+A(r) \;dr^2+r^2 d\theta^2+r^2 \sin^2\theta d\phi^2\;.
\end{eqnarray}
The action (\ref{action}) yields two independent field equations due to the Bianchi identity. For asymptotically flat solutions, $B(r)$ is generally set to unity at infinity by convention.

The metric behaviour can be expressed by the series expansion of the metric functions around $r=0$~\cite{Stelle:1977ry} as
\begin{eqnarray}
A(r)&=&a_s r^s+a_{s+1}r^{s+1}+a_{s+2}r^{s+2}+...\;,\nonumber\\
B(r)&=& b_t (r^t+b_{t+1}r^{t+1}+b_{t+2}r^{t+2}+...)\;.
\end{eqnarray}
where each solution is identified by the powers of the first non-vanishing terms, $(s,t)$. The action (\ref{action}) yields three families of solutions; $(0,0),\;(1,-1),\;(2,2)$. The $(0,0)$ family corresponds to a star-like structure and could substantially differ from their GR counterparts~\cite{Holdom:2016nek}. The $(1,-1)$ family represents black-hole-like solutions, where the Schwarzschild solution appears as a special case, which is not surprising since the vacuum solutions of GR automatically satisfy the vacuum field equations generated by the action of quadratic gravity~(\ref{action}). The $(2,2)$ solutions, i.e. 2-2-holes, are characterised by a metric vanishing at the origin and do not have an analog in GR. In Ref.~\cite{Holdom:2016nek}, the basic properties of these objects (for a particular class of solutions which have only even-power terms in the series expansion above) were studied in a case where they are sourced by matter, in a simple thin-shell toy model in the large-mass limit. 2-2-hole solutions, in addition to black holes, are found when the shell radius is smaller than would-be horizon, where the star-like (0,0) solution doesn't exist as in the case of GR. Therefore, 2-2-holes can serve as alternatives to black holes for the endpoint of gravitational collapse. 

A 2-2-hole, being well-behaved everywhere except the origin, does not posses an event horizon. It resembles closely the Schwarzschild solution in the exterior and begins to diverge close to the would-be horizon, whereas in the interior a high-curvature novel behaviour dominated by the quadratic curvature terms takes over. There exists a timelike singularity at the origin, which can be interpreted as an indication for geodesic incompleteness of the spacetime. It would not be inconceivable to think that this concerns only the motion of classical point particles. A possible way to probe this singularity with energy packets in a relativistic classical field theory was discussed in~\cite{Holdom:2016nek}.

Recently, the investigation continued in a more realistic picture in Refs.~\cite{Holdom:2019ouz,Ren:2019afg}, in which 2-2-hole solutions sourced by a thermal gas were found and explored in detail. It turns out that the thermal gas in quadratic gravity is able to survive as an ultracompact configuration without collapsing into a black hole, unlike in the case of GR~\cite{Sorkin:1981wd}. The thermal-gas model provides a useful scheme for understanding the thermodynamical behaviour of 2-2-holes and for physical applications such as addressing the dark matter problem. 

For the thermal 2-2-holes, we need the energy-momentum tensor for the thermal gas which is given as
\begin{eqnarray}
\label{stresstensor}
T_{\mu\nu}&=&\mathrm{diag} (B\rho, A p, r^2 p, r^2 p \sin^2\theta ),\nonumber\\
\rho&=&\frac{g}{(2\pi)^3}\int^\infty_0\frac{E}{e^{E/T}-\epsilon}4\pi \mathrm{p}^2 d\mathrm{p}\nonumber\\
p&=&\frac{g}{3(2\pi)^3}\int^\infty_0\frac{\mathrm{p}^2/E}{e^{E/T}-\epsilon}4\pi \mathrm{p}^2 d\mathrm{p}
\end{eqnarray}
where $E^2=m^2+\mathrm{p}^2$, and $T$ is the local temperature. The symbols $\rho$, $p$, and $\mathrm{p}$ denote the proper energy density, the isotropic pressure, and the 3-momentum, respectively. $g$ is the number of degrees of freedom for each species, and $\epsilon = \pm1$ for bosons and fermions. Note that the convention $c=\hbar=k_B=1$ is adopted throughout this work, unless stated otherwise.
\begin{figure}[!h]
\hspace{-1.2cm}
{ \includegraphics[width=8.0cm]{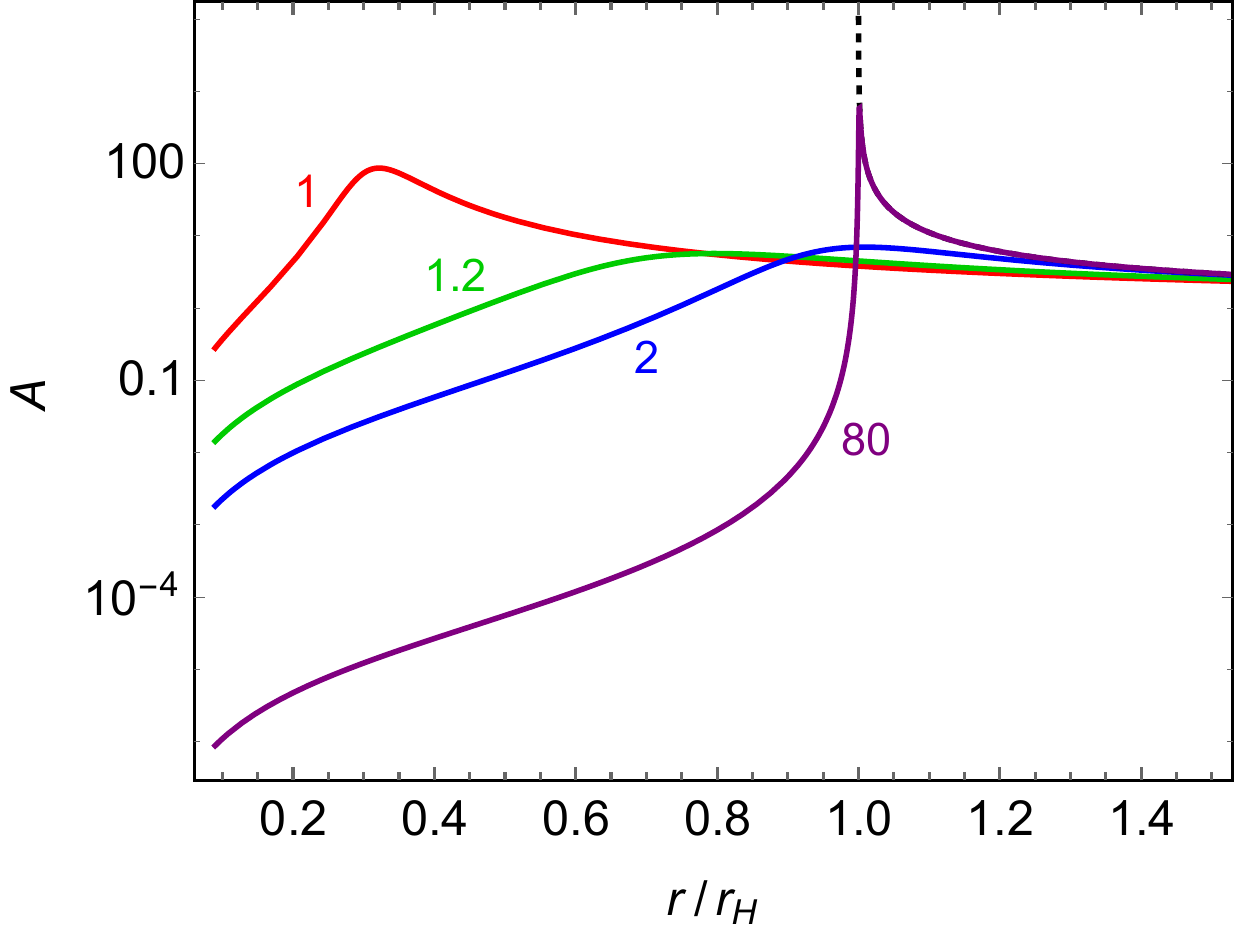}}\quad
{ \includegraphics[width=8.0cm]{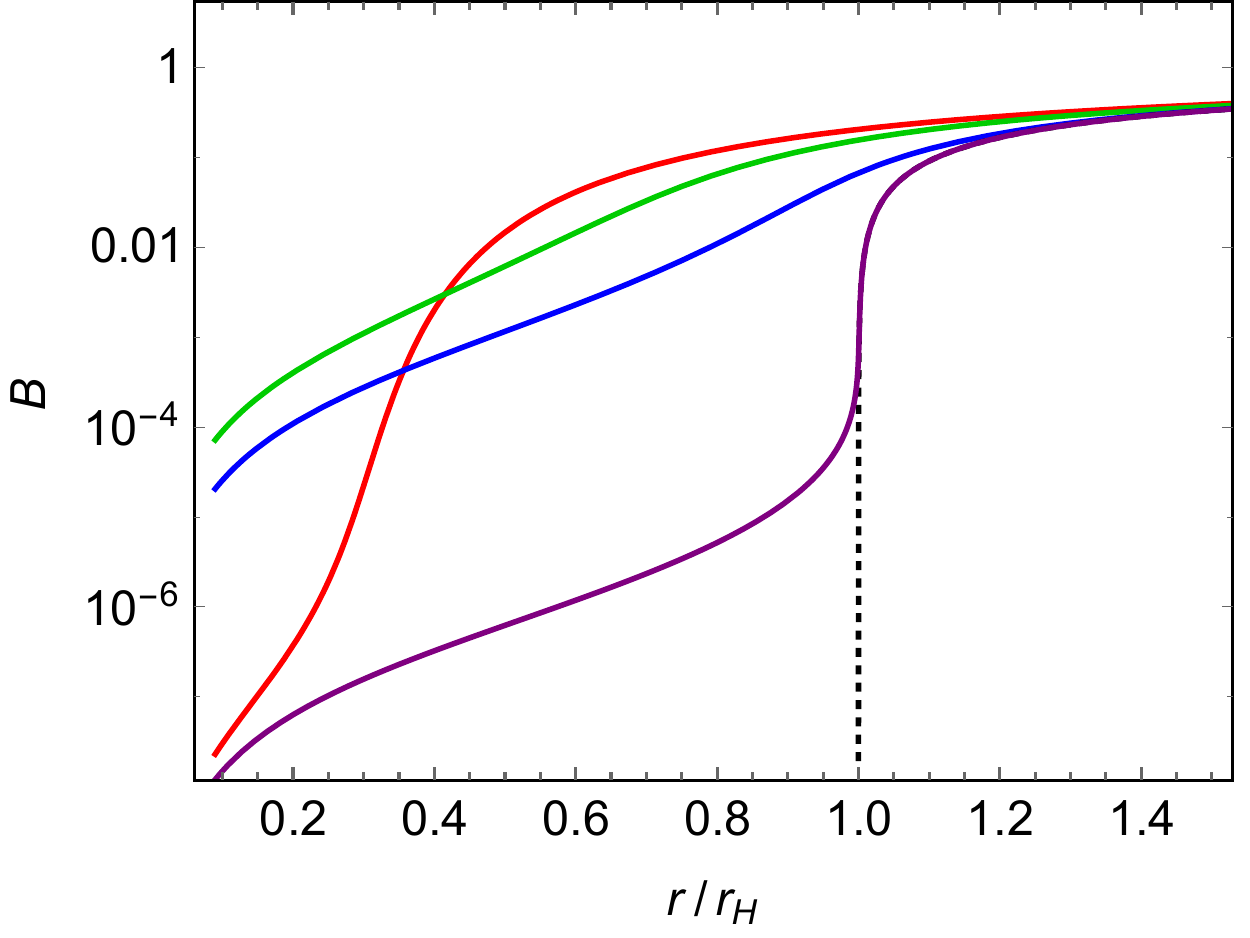}}
\caption{\label{fig:solutions} 
A snapshot of solutions for thermal 2-2-holes~\cite{Holdom:2019ouz,Ren:2019afg}. The metric functions $A(r)$ and $B(r)$ are displayed for $M/\Mmin\approx 1$ (red), 1.2 (green), 2 (blue), 80 (purple). The black dotted line is the Schwarzschild solution. Taken from~\cite{Aydemir:2020xfd}.}
\end{figure}

The existence of 2-2-hole solutions relies on the Weyl term $C^{\mu\nu\rho\sigma}C_{\mu\nu\rho\sigma}$ in the quadratic action, whereas the $R^2$ term is optional and therefore is neglected for simplicity. The Weyl term introduces a new spin-2 mode with mass $m_2$, which determines the minimum mass for the 2-2-hole as
\begin{eqnarray}
\label{eq:Mmin}
\hMmin\equiv\frac{\Mmin}{\Mp}\approx 0.63\frac{\Mp}{ m_2}\approx 0.63 \frac{\lambda_2}{\lp}.
\end{eqnarray}
This indicates that the size of 2-2-holes is bounded from below by the corresponding Compton wavelength $\lambda_2$.
Depending on the strength of dimensionless couplings associated with the quadratic curvature terms, there exist two scenarios for quadratic gravity. In the strong coupling scenario, the Planck mass emerges dynamically through dimensional transmutation as the only mass scale, $m_2\approx \Mp$, i.e. $\hMmin\approx 0.63$. In the weak coupling scenario, the Planck mass can be generated either spontaneously through vacuum expectation values of some scalar fields or it can be introduced explicitly. For the weak coupling case, there can be a large mass-hierarchy with $m_2\ll \Mp$, i.e. $\hMmin\gg 1$. 

The solutions for spherically symmetric, asymptotically flat and static case~\cite{Holdom:2019ouz,Ren:2019afg}, which can only be found numerically due to the nontrivial field equations, are shown in Fig.~\ref{fig:solutions} for various values of $M/\Mmin$. A 2-2-hole, as in the case of thin-shell model previously mentioned, possesses an exterior that resembles the Schwarzschild solution with the same physical mass $M$, an interior characterized by a high-curvature region, and a transition region around the would-be horizon $r_H=2M\lp^2$ connecting the two sides. A large hole with $M\gtrsim M_\textrm{peak}=1.2\Mmin$ has an extremely narrow transition region around the would-be horizon $r_H$, and it appears very much like a black hole for an outside observer.
In contrast, a small 2-2-hole with $M\lesssim M_\textrm{peak}$ ("small" refers to $M$ being close to $\Mmin$ even when $\Mmin$ is large), has a broader transition region and a shrinking interior.


\subsection{Thermodynamics and evaporation of thermal 2-2-holes}
The total energy and entropy of the thermal gas are respectively given as
\begin{eqnarray}
U&=&\int dr \sqrt{A(r)} \;4\pi r^2\; \left(\sqrt{B(r)}\;\rho (r)\right)\nonumber\\
S&=& \int dr \sqrt{A(r)}\; 4\pi r^2\; s (r)\;,
\end{eqnarray}
where $s(r)\equiv (\rho(r)+p(r))/T(r)$. The proper volume element and the redshift effects are taken into account in the expression above, based on the metric (\ref{metric}). Focusing on the relativistic case and hence ignoring the mass of the species, we obtain from (\ref{stresstensor})
\begin{equation}
\label{rho}
\rho=3p=\frac{\pi^2}{30} N\, T^{4}\,.
\end{equation}
$T(r)$ is the local measured temperature and 
$N= g_b+7 g_f/8$. In principle, $N$ includes particle species of all kinds and could be much larger than the  Standard Model value ($N\approx 107$). As a results of the conservation law of the stress tensor, the local temperature $T(r)$ satisfies Tolman's law ($T(r)g_{00}^{1/2}=T_\infty$), where $T_\infty$ is the value measured by a distant observer and represents the temperature at which the 2-2-hole radiates as black body when it is not in thermal equilibrium with its surroundings. The relation between the total energy and the entropy of the gas is given as
\begin{eqnarray}
S=\frac{4}{3}\frac{U}{T_\infty}=\frac{8\pi^3}{45} N \,T^3_\infty\int dr \sqrt\frac{A(r)}{B(r)^3}r^2\,.
\end{eqnarray}
The thermal 2-2-holes exhibit intriguing thermodynamic behavior for the small- and large-mass cases, as discussed below.

 In the large-mass range, the temperature and entropy of the 2-2-hole of $M$ are well approximated by the relations~\cite{Ren:2019afg}
\begin{eqnarray}\label{eq:LMlimit}
T_\infty\approx 1.7\, N^{-1/4}\hMmin^{1/2}\, T_\textrm{BH},\quad
S\approx 0.60\, N^{1/4}\hMmin^{-1/2} \,S_\textrm{BH}\,,
\end{eqnarray}
where $T_\textrm{BH}=\Mp^2/8\pi M$ is the Hawking temperature and $S_\textrm{BH}=\pi\, r_H^2/\lp^2$ is the Bekenstein-Hawking entropy for a Schwarzschild black hole with the same mass. 
In other words, a 2-2-hole in the early radiation stage exhibits the anomalous behavior of black hole thermodynamics, i.e. the negative heat capacity and the area law for entropy, which in this case arises from the ordinary thermal gas on a highly curved background spacetime. Consequently, a large 2-2-hole appears similar to a black hole for an outside observer as far as its thermodynamic behavior is concerned. It deviates from a black hole due to the dependence of its thermodynamic quantities on the number of degrees of freedom $N$ and the minimal mass $\Mmin$. 
In the strong coupling scenario, where $\hMmin\approx 1$, the difference mainly comes from the $N$ dependence. For a reasonable choice of $N$, e.g. the Standard Model value, the 2-2-hole entropy can be larger than the entropy of the black hole with the same mass. This suggests that a 2-2-hole is thermodynamically more stable, and hence would be favored as the endpoint of gravitational collapse. In the weak coupling scenario ($\hMmin\gg 1$), on the other hand, 2-2-holes have much higher temperature and much smaller entropy. Thus in this case, 2-2-holes are no longer entropically favorable and their stability needs to be examined dynamically. 
Despite of the difference in thermodynamic quantities, (\ref{eq:LMlimit}) still satisfies $T_\infty S= T_\textrm{BH} S_\textrm{BH}=M/2$, which is consistent with the first law of thermodynamics. This also determines the total energy $U=3M/8$ for the gas, meaning that a sizable fraction of the physical mass for the hole comes from the gas source. 


 In the small-mass range, the temperature and entropy are obtained at the leading-order as~\cite{Ren:2019afg}
\begin{eqnarray}\label{eq:SMlimit}
T_\infty\approx 0.39\, N^{-1/4}\hMmin^{-3/2}\Delta M\left(\ln\frac{M_\textrm{min}}{\Delta M}\right)^{7/4},\,\,
S\approx 3.4\, N^{1/4}\hMmin^{3/2}\left(\ln\frac{M_\textrm{min}}{\Delta M}\right)^{-3/4}.
\end{eqnarray}
A 2-2-hole in this stage has a positive heat capacity and behaves like a classical thermodynamic system. In the limit $\Delta M\to 0$, the temperature approaches zero almost linearly in $\Delta M$, whereas the entropy decreases much slower. Then, the energy in this limit is dominated by the gravitational field, and the contribution from the gas becomes negligible. 

Although the exact solutions could only be found numerically, with the large-mass analytical approximation (\ref{eq:LMlimit}) applied to $M\gtrsim \Mpeak$ and the small-mass one (\ref{eq:SMlimit}) applied to $M\lesssim \Mpeak$, the whole mass range is accurately described, including the estimation in the intermediate region around the temperature peak with~\cite{Ren:2019afg,Aydemir:2020xfd}
\begin{eqnarray}
\label{eq:peak}
T_{\infty,\textrm{peak}}\approx 0.050\,\Mp\, N^{-1/4}\hMmin^{-1/2}\,\,
\textrm{ at }\,\, \Mpeak\approx 1.2 M_\textrm{min}\,. 
\end{eqnarray}

Now, we proceed to the evaporation of 2-2-holes. If a thermal 2-2-hole is hotter than the cosmic microwave background, it will radiate.  The mass evolution is obtained through the power formula
\begin{eqnarray}
P=-\frac{dM}{dt}=\int^{\infty}_0 E \frac{dN}{dt dE} dE\;,
\end{eqnarray}
where the emission rate is given as
\begin{eqnarray}
\label{emissionrate0}
\frac{dN}{dt dE} = \frac{1}{2\pi^2}\frac{E^2 \sigma_a (M,E)}{e^{E/T_{\infty}}-(-1)^{2s}}\;.
\end{eqnarray}
Here, $s$ denotes the spin of the emitted particle.  $\sigma_a (M,E)$ is the absorption cross section and should be determined for each species separately, as in the case of black holes~\cite{MacGibbon:1991tj,Carr:2009jm}. For convenience, we use $\sigma_a (M,E)=\pi r_H^2$ that makes a good approximation, in overall. This lead to the common version of the Stefan-Boltzmann power formula
\begin{equation}\label{eq:SBlaw}
-\frac{dM}{dt}
\approx \frac{\pi^2}{120}\, N_* \, A \,T_{\infty}^4\;,
\end{equation}
which assumes $A=4\pi r_H^2$ as the effective emitted area. $N_*$ is in the same form as $N$, defined in (\ref{rho}), but unlike $N$ it accounts for, not the total existing degrees for freedom corresponding to the thermal gas in the interior, but the radiation degrees of freedom, and changes depending on the temperature of interest. It includes particles lighter than $T_{\infty}$ and it can be much smaller than $N$. For simplicity, we treat $N_*$ as a constant, determined by the initial $T_{\infty}$. 
In the Standard Model, $N_*\approx 107$ for $T_{\infty}\gtrsim\,$TeV, $N_*\approx 62$ for $T_{\infty} \sim$\,GeV, and $N_*\approx 11$ for $T_{\infty}\sim$\,MeV.


In the large-mass stage, the time dependences of the temperature and mass take the same form as a black hole, 
\begin{eqnarray}\label{eq:LMlimitTime0}
T_\infty(t)\approx T_{\infty,\textrm{init}}\left(1-\frac{\Delta t}{\tau_L}\right)^{-1/3},\quad
M(t)\approx \Mini\left(1-\frac{\Delta t}{\tau_L}\right)^{1/3}\;.
\end{eqnarray}
as can be found from (\ref{eq:LMlimit}) and (\ref{eq:SBlaw}). Here, $\Delta t \equiv t-t_\textrm{init}$ is the time it takes for a 2-2-hole to evolve from $\Mini$ to $M\gtrsim \Mpeak$, whereas the time spent in the large-mass stage in total is obtained as
\begin{eqnarray}\label{eq:tauL}
\tau_L\equiv t_\textrm{peak}-t_\textrm{init}
\,\approx \,N\,N_*^{-1}  \hMmin^{-2} \left(\frac{\Mini}{3.7\times 10^8\,\textrm{g}}\right)^3\textrm{s}\,
\end{eqnarray}
which can also be written in terms of $T_{\infty,\textrm{init}}$, instead of $\Mini$, from (\ref{eq:LMlimit}).
 Notice that, in comparison to a black hole with the lifetime $\tau_\textrm{BH}=\tau_L$, the expressions above differ only by an overall constant, as expected. Substituting $\Mini$ and $T_{\infty,\textrm{init}}$ as functions of $\tau_L$, (\ref{eq:LMlimitTime0}) becomes
\begin{eqnarray}\label{eq:LMlimitTime}
T_\infty(t)&\approx& 1.1\,\Mp\, N^{1/12}N_*^{-1/3} \hMmin^{-1/6}\left(\frac{\tau_L-\Delta t}{\lp}\right)^{-1/3},\nonumber\\
M(t)&\approx& 0.064\,\Mp \,N^{-1/3}N_*^{1/3}\hMmin^{2/3}\left(\frac{\tau_L-\Delta t}{\lp}\right)^{1/3},
\end{eqnarray}
where we can see the explicit $\Mmin$ dependence. 

 In the small-mass stage, by using (\ref{eq:SMlimit}) and (\ref{eq:SBlaw}), we obtain 
\begin{eqnarray}\label{eq:SMlimitTime}
T_\infty(t)\approx 1.1\,\Mp\,  N^{1/12} N_*^{-1/3} \hMmin^{-1/6} \left(\frac{\Delta t-\tau_L}{\lp}\right)^{-1/3}\left(\ln\frac{\Delta t-\tau_L}{\lp\,\hMmin}\right)^{-7/12}\;,
\end{eqnarray}
at the leading order.
The time dependence of the mass difference $\Delta M(t)$ can be found from (\ref{eq:SMlimit}) and (\ref{eq:SMlimitTime}) as
\begin{eqnarray}
\Delta M(t)\approx 19\, \Mp\, N^{1/3} N_*^{-1/3} \hat{M}_{\mathrm{min}}^{4/3}\left(\frac{\Delta t-\tau_L}{\lp}\right)^{-1/3}\left(\ln \frac{\Delta t-\tau_L}{\lp\,\hat{M}_{\mathrm{min}}} \right)^{-7/3}\;,
\end{eqnarray}  
which is extremely small, meaning the mass in this stage is close to $\Mmin$.

\subsection{Remarks on the entropy-area law}

Thermodynamical entropy is an extensive quantity, thus it is generally expected to scale with the volume of the system.
 One of the mysterious features of black hole thermodynamics is that the entropy is given by the horizon area. As well-known, from the standpoint of classical physics, since there is no radiation escaping from black holes, it has a vanishing temperature and the concept of entropy becomes meaningless. However, once the particle creation effects on the curved background
 are taken into account in the semiclassical treatment of gravity, it turns out that a black hole thermally radiates with the Hawking temperature~\cite{Hawking:1974sw}. The entropy can be inferred from the first law of thermodynamics as the Bekenstein-Hawking relation $S_\textrm{BH}=A/4\lp^2$, which also has been confirmed through various derivations~\cite{Gibbons:1976ue,Banados:1993qp,Wald:1993nt,Iyer:1995kg}. So, the relations first considered as analogies between the black hole mechanics and thermodynamics~\cite{Bekenstein:1973ur,Bekenstein:1974ax} come out as well-established true thermodynamical effects.



It has been argued in the literature that the entropy-area law might be a generic feature in Nature when the quantum behaviour is taken into account.  For instance, it was shown in~\cite{Srednicki:1993im} (see also~\cite{Bombelli:1986rw}) that when the ground state density matrix for a massless free scalar field on a Minkowski background is traced over the degrees of freedom within an "inaccessible" spherical region (which does not have to be causally disconnected from the outside region), the von Neumann entropy (associated with the reduced density matrix $\rho_{\textrm{out}}$) $S=-\textrm{tr}(\rho_{\textrm{out}}\ln\rho_{\textrm{out}})$, i.e. entanglement entropy, is proportional the surface area. Here, the entropy can be interpreted as the lack of information on the traced-over states. The entropy-area law has been studied in many contexts~\cite{Eisert:2008ur,tHooft:1993dmi,Susskind:1994vu} and has crucial implications regarding information transfer regardless of horizon formation, but particularly for black holes in the context of the information-loss problem due to the existence of horizon~\cite{Chen:2014jwq}. Yet the entropy-area law in black holes is not generic when theories beyond GR are considered~\cite{Wald:1993nt,Iyer:1995kg}. 

The situation becomes even more interesting in the case of 2-2-holes. As previously mentioned, a small 2-2-hole, in the late-time-radiation (or remnant) stage, behaves like an ordinary thermodynamic system, and the temperature and entropy, with logarithmic dependences in the leading order, are in quite different forms than the black-hole case. However, for large 2-2-holes, the thermodynamic behaviour is similar to that of a BH, with $T_{\infty}\propto T_\textrm{BH} $ and the entropy satisfying the area law $S\propto S_\textrm{BH}$. In contrast to black holes, in the case of large (thermal) 2-2-holes, this behaviour \textit{explicitly} arises from self-gravitating relativistic thermal gas on a curved background, without taking into account spontaneous particle creation from vacuum.\footnote{Taking into account particle-creation effects from vacuum on the curved background for 2-2-holes (as in the case of black-holes) would introduce an extra contribution expected to be on the order of  $T_\textrm{BH}$ and $S_\textrm{BH}$ to the temperature and entropy, respectively.}
~The resulting similarity between the black-hole thermodynamics and that of a (large) 2-2-hole, notwithstanding their different origins, is quite remarkable. 
This requires further investigation and could provide insights on the relation between classical and quantum gravity through thermodynamics, as well as on the role of gravity in information transfer and entanglement entropy.

\section{Primordial 2-2-holes remnants as dark matter}
\label{sec:dmintro}

 Primordial 2-2-holes, just like PBHs, can form in the early universe when some regions of the universe stop expanding and re-collapse. This could occur mainly due to density inhomogeneities seeded by inflation or due to a first order phase transition, where the former is assumed in this paper. The formation is generally considered in the radiation era, although it is possible that it occurs in a transient matter-domination period before the matter-radiation equality~\cite{Dalianis:2019asr}.
 

We focus on the generic case that  the background is radiation dominated at the time of formation. 
The initial mass $\Mini$ can be no larger than the horizon mass at the formation time ($t_\textrm{init}$), $\Mp^2/2H(t_\textrm{init})\approx 4\times 10^{38}\left(t_\textrm{init}/\textrm{s}\right)\,\textrm{g}$.
Assuming that the reheating temperature is no larger than $10^{16}\,$GeV~\cite{Dalianis:2019asr}, the minimum horizon mass is given as $\sim 1\,$g at the end of inflation, and the maximum horizon mass is given as $\sim 10^{50}$\,g at matter-radiation equality.

 A typical 2-2-hole is expected to be formed with the initial mass $\Mini$ much larger than $\Mmin$.
 The phenomenology then strongly depends on the duration of the early stage of evaporation $\tau_L$,  as given in (\ref{eq:tauL}), determined by $\Mini$ for a given $\Mmin$. For later discussion, it is convenient to define the following critical masses for $\Mini$,
\begin{eqnarray}\label{eq:Mi}
\left(M_\textrm{uni},\, M_\textrm{rec},\, M_\textrm{BBN}\right)
=\left(2.8\times 10^{14},\, 8.8\times 10^{12},\,3.7\times 10^8\right)\hMmin^{2/3}\;N^{-1/3}\,N_*^{1/3} \;\textrm{g} \,,
\end{eqnarray} 
corresponding to $\tau_L\approx t_0\approx 4.3 \times 10^{17}\,$s (the age of the universe), $10^{13}\,$s (recombination), $1\,$s (BBN). 
Note that in the strong coupling scenario the mass values above are comparable to those for PBHs, while they can be much larger in the weak coupling scenario given the $\hMmin^{2/3}$ dependence. So for the ones that survive until today we have  $\Mini\gtrsim 10^{14}$. 
We focus on the 2-2-holes that have become remnant by today, meaning the mass range of interest is $\Mini\lesssim M_\textrm{uni}$.

The remnant mass $\Mmin$, which is determined by the fundamental mass scale $m_2$ in the theory, can be probed by the present epoch observations. $\Mmin$ has a theoretical lower bound, $\Mmin\gtrsim 0.63\,\Mp$, corresponding to  the strong coupling scenario.  The solar system tests of GR provide a rough upper bound $\Mmin\lesssim 10^{33}\,\textrm{g}\sim M_\odot$, by the requirement that the Compton wavelength $\lambda_2$ be no larger than $\mathcal{O}(\textrm{km})$. An isolated remnant, in the same way as a PHB, could be detected through its gravitational interaction. Note the relevant studies generally assume a Newtonian force for the object, so in that case a 2-2-hole remnant that deviates at $r\sim \mathcal{O}(r_H)$ would still appear indistinguishable from a black hole.
The constrained parameter space covers the range $\Mmin\gtrsim 10^{17}$\,g, with some examples summarized in Fig.~\ref{fig:LUcons}. Thus, only smaller remnants, with relatively weak gravitational interactions, can account for the entirety of dark matter.    

The mass fraction at formation in the radiation era is given as
\begin{eqnarray}\label{eq:beta}
\beta \equiv \frac{\rho(t_\textrm{init})}{\rho_\textrm{tot}(t_\textrm{init})}
=\frac{4 \,M(t_\textrm{init})\,n(t_\textrm{init})}{3 \,T(t_\textrm{init})\,s(t_\textrm{init})}
=2.5\, g_*^{1/4}\,\gamma^{-1/2}\,\hMini^{3/2}\,\frac{n(t_\textrm{init})}{s(t_\textrm{init})}\,,
\end{eqnarray}
where $\hMini\equiv \Mini/\Mp$. $\rho(t), \,n(t)$ denote the energy density and number density for the objects of interest. $ g_*$  counts the relativistic background degrees of freedom at the corresponding temperature. The expression above is valid for any primordial object such as a 2-2-hole or a black hole. 
$\gamma$ denotes the fraction of the horizon mass consumed for the 2-2-hole formation. A naive analytical calculation suggests $\gamma\approx 0.2$, but this is highly uncertain~\cite{Carr:2009jm}. 
The observational constraints can be expressed in terms of the number density to entropy density ratio $n(t_\textrm{init})/s(t_\textrm{init})$, namely, the combination $\beta\,\gamma^{1/2}g_*^{-1/4}$, which is insensitive to $\gamma$. 

The mass fraction of primordial objects in dark matter at the present epoch is,
\begin{eqnarray}\label{eq:betaf0}
f=\frac{M(t_0)\,n(t_0)}{\rho_\textrm{DM}(t_0)}
=\frac{M(t_0)\,s(t_0)}{\rho_\textrm{DM}(t_0)}\frac{n(t_0)}{s(t_0)}\,,
\end{eqnarray} 
where $\rho_{\textrm{DM}}(t_0)\approx 0.26\rho_c$, $\rho_c=9.5\times10^{-30}\,\textrm{g}\,\textrm{cm}^{-3}$, and $s(t_0)=2.9\times 10^3\,\textrm{cm}^{-3}$. For the large-mass case, where the evaporating rate is negligible,  $M(t_0)$ is $\Mini$. For the small-mass case, on the other hand, $M(t_0)$ is $\Mmin$, which is the mass of the remnant left behind. Note that here we approximate the evaporation as 
an instantaneous radiation of energy at $t_\textrm{eva}\equiv t_\textrm{init}+\tau_L\approx \tau_L$ (or $\tau_\textrm{BH}$). This can be considered as a good approximation, given that $M(t)$ varies slowly with time as compared to other quantities at both early and late times.


\section{Phenomenology}
\label{sec:pheno0}
In this section, we discuss observational implications of having 2-2-hole remnants as dark matter, and the corresponding constraints from the present-epoch observations and the early-universe cosmology.

The remnants, in principle, can be probed through the possible thermal radiation they emit. However, the amount produced from isolated 2-2-hole remnants is expected to be relatively weak. A conservative estimation shows that the contribution to the diffuse photon flux at present is insignificant, rendering them difficult to be detected through this way.
On the other hand, if two remnants form a binary and merge, then the merger product is not a the remnant state anymore but a hot 2-2-hole, which produces a large amount of radiation in a very short time-period. The corresponding experimental constraints from binary mergers turn out to be significant, providing a new testing opportunity for small size dark matter that only interacts gravitationally with normal matter. 

Early universe constraints, on the other hand, require that 2-2-holes reside in the remnant state before BBN.  A conservative estimation shows that the remnant radiation with the dark matter abundance can safely evade BBN and CMB constraints. 

\subsection{The present-epoch signatures}

\textit{High energy particle flux from binary mergers:}
A 2-2-hole remnant can be pushed away from the remnant stage if it is able to absorb sufficient mass. Although the merger of two 2-2-hole remnants and the accretion of ordinary matter onto a remnant can both contribute, the former is the more likely mechanism. 

 When a binary of two remnants merge, the merger product is no longer a remnant state, but a hot 2-2-hole with $M_\textrm{merger}\approx 2\Mmin>\Mpeak$.
Then, from (\ref{eq:LMlimit}), the temperature of the merger product is 
\begin{eqnarray}
\label{Temp-merger}
T_{\infty,\textrm{merger}}=3.4\times 10^{-2}\,\Mp\,N^{-1/4}\hMmin^{-1/2}
=1.9\times10^{15} N^{-1/4}\left(\frac{\Mmin}{\textrm{g}}\right)^{-1/2}\,\textrm{GeV}\,.
\end{eqnarray}
The average energy of emitted particles, which can be approximated as the temperature, drops as $\Mmin$ increases and can be significantly high for small $\Mmin$.  For a Planck-mass remnant, the particles could have roughly the Planck energy, whereas for a large remnant with $\Mmin\sim 10^{23}\,$g, the energy is around TeV scale. With the lifetime in this early radiation stage ($\tau_L$) being much smaller than a second for this mass range, it is assured that such mergers release their excess energy almost instantly to get back to the remnant phase, producing fluxes of high energy particles. Therefore, observations in high-energy astrophysical particle searches can be used to probe 2-2-hole remnants and constraint the remnant mass $\Mmin$, a fundamental scale in the theory.

Ultra-high energy cosmic rays with energies up to ($\gtrsim10^{11}\,$GeV) have long been observed. The lack of suppression in the observed flux beyond the Greisen-Zatsepin-Kuzmin (GZK) cut-off ($\sim5\times 10^{11}\,$GeV) reported in late 90's stimulated much interest in search for a new-physics explanation.  However, a clear suppression around $10^{11}\,$GeV is now seen in the latest observations~\cite{ThePierreAuger:2013eja,Abu-Zayyad:2013qwa}, thus the need for new physics is no longer as strongly motivated. Yet, these signals remain as mystery since their sources are still unknown. In recent years, the photon flux around the same energy has also been measured with improved precision \cite{Aab:2016agp,Abbasi:2018ywn}. High-energy neutrino experiments, on the other hand, probe a much wider energy range from $10^{3}\,$GeV to $10^{16}\,$GeV~\cite{Buitink:2010qn,Gorham:2019guw,Gorham:2010kv,Aartsen:2018vtx,Aartsen:2015xup,Aartsen:2015knd,Aartsen:2014qna}. We use these observations to constraint our parameter space.

Since the processes we are dealing with are high energy emissions, the dominant contribution to flux comes from high-multiplicity final states with a broad energy spectrum rather than a small number of particles with energies identified by the temperature. These high-multiplicity states emanate from generalized parton showers of highly off-shell initial particles. A parton shower begins with the fragmentation of initial quarks or gluons into nucleons, then results in generation of photons and neutrinos from decay of hadrons. Note that showers can also be initiated by the initial particles with only the electroweak charges depending on the strength of relevant couplings. Here, we will consider showers initiated by quarks only, since they are the dominant product of the instantaneous 2-2-hole evaporation, in order to compute the flux of the secondary production of protons, neutrinos, and photons. In case of neutrinos we will also look at the contribution from the primary production. As mentioned above, neutrinos as the initial particles can also instigate a shower, but the shower spectrum of neutrinos peaks around the maximum energies~\cite{Barbot:2002gt}, implying weak coupling and less showering, and therefore it is a reasonable approximation to ignore the showering initiated by neutrinos.\footnote{In case of photons; the primary production, which would approximately identify the particle energies with the temperature, would not be useful for the 2-2-hole mass-range of interest since there do not exist photon bounds to be compared at such high energies. On the other hand, showers initiated by the photons would in fact give contributions at low energies due to the energy spectrum of final particles, but the overall effect would be sub-leading to the contributions from the showers instigated by quarks. Therefore, we ignore the cases where photons are the initial particles.} Additionally, for neutrinos we should also include the extragalactic contribution as opposed to protons and photons whose contributions are suppressed due to their interaction with CMB photons at such high energies.
 
We start with the galactic contribution to the flux from on-shell neutrinos. The neutrino flux from binary mergers of 2-2-hole remnants   can be estimated as 
\begin{eqnarray}
\label{flux}
\Phi_{\nu}=\frac{D}{2 M_{\textrm{min}}}\,\frac{d N_{\nu}}{d E_\nu dt}\,.
\end{eqnarray}
where $D$ is the relevant astrophysical factor~\cite{Evans:2016xwx} for the Milky Way with Einasto density profile~\cite{Pato:2015dua}. The neutrino emission rate is given as
\begin{eqnarray}
\frac{d N_{\nu}}{d E_\nu dt}\approx \eta_{\nu} \frac{M_{\textrm{min}}}{\langle{E}_{\nu}\rangle^2}\;\Gamma\,,
\end{eqnarray}
as the discharge of the excess energy, in the amount of one remnant mass. We approximate the spectrum by on-shell emission at the average energy $\langle{E}_{\nu}\rangle\approx 4.2$ $\,T_{\infty,\textrm{merger}}$. Here, $\eta_{\nu}\approx 0.058$ denotes the fraction of the total energy as neutrinos in case of the particle content of the Standard Model~\cite{MacGibbon:1991tj}. $\Gamma$ is the merger rate of 2-2-hole remnants. It was recently suggested in \cite{Raidal:2018bbj,Vaskonen:2019jpv} that the merger rate might be suppressed compared to the earlier estimation, once disruptions of the binaries from nearby holes are taken into account.
In the case of disruption, the dominant contribution is received from the non-perturbed binaries $\Gamma\approx\Gamma_{\textrm{np}}\textrm{P}_{\textrm{np}} $, where  $\textrm{P}_{\textrm{np}}$ is the fraction of binaries remaining unperturbed. $\Gamma_{\textrm{np}}$ can be obtained from the total rate per volume given in \cite{Vaskonen:2019jpv} as
\begin{eqnarray}
\Gamma_{\textrm{np}} &=&4.7\times 10^{-26}\left(1+5.8\times 10^{-5}f^{-2}\right)^{-21/74}\,f^{16/37}\left(\frac{\Mmin}{\textrm{g}}\right)^{5/37} \left(\frac{t}{t_0}\right)^{-34/37} \textrm{s}^{-1}\;.
\label{mergerrates}
\end{eqnarray}
 The functional form $\textrm{P}_{\textrm{np}}$, which is relevant for $f\gtrsim 4\times 10^{-3}$, can be approximated as  $\textrm{P}_{\textrm{np}}\approx8.2\times10^{-3} f^{-4/5}$. For smaller $f$, the disruption effects are not effective, so the no-disruption case is recovered with $\textrm{P}_{\textrm{np}}\approx 1$. 

The flux contribution from showers of the highly off-shell initial particles is taken into account by multiplying the emission rate of the initial particles, with the fragmentation function $D_q^j(x)$, $j=p,\,\gamma,\,\nu$. Note that as an approximation we consider only quarks as the particles that initiated the showers since they are much more abundant than gluons in the emission spectrum of the 2-2-hole. That is to say, in analogy with (\ref{flux}), we have
\begin{eqnarray}
\Phi_{j}=\frac{D}{2 M_{\textrm{min}}}\left[\frac{d N_{q}}{d E_q dt}\right]\Big[D_q^{j}(x)\Big]_{x=E_j/E_q}
\end{eqnarray}
for the galactic contribution, where $\langle{E_q}\rangle\approx 4.2\,T_{\infty,\textrm{merger}}$ and $\eta_q\approx 0.67$. We obtain the fragmentation functions $D_q^j(x)$ for ordinary QCD for such high energies by utilising the results in the literature~\cite{Aloisio:2003xj,Barbot:2002gt}.

\begin{figure}[!h]
  \centering%
{ \includegraphics[width=11cm]{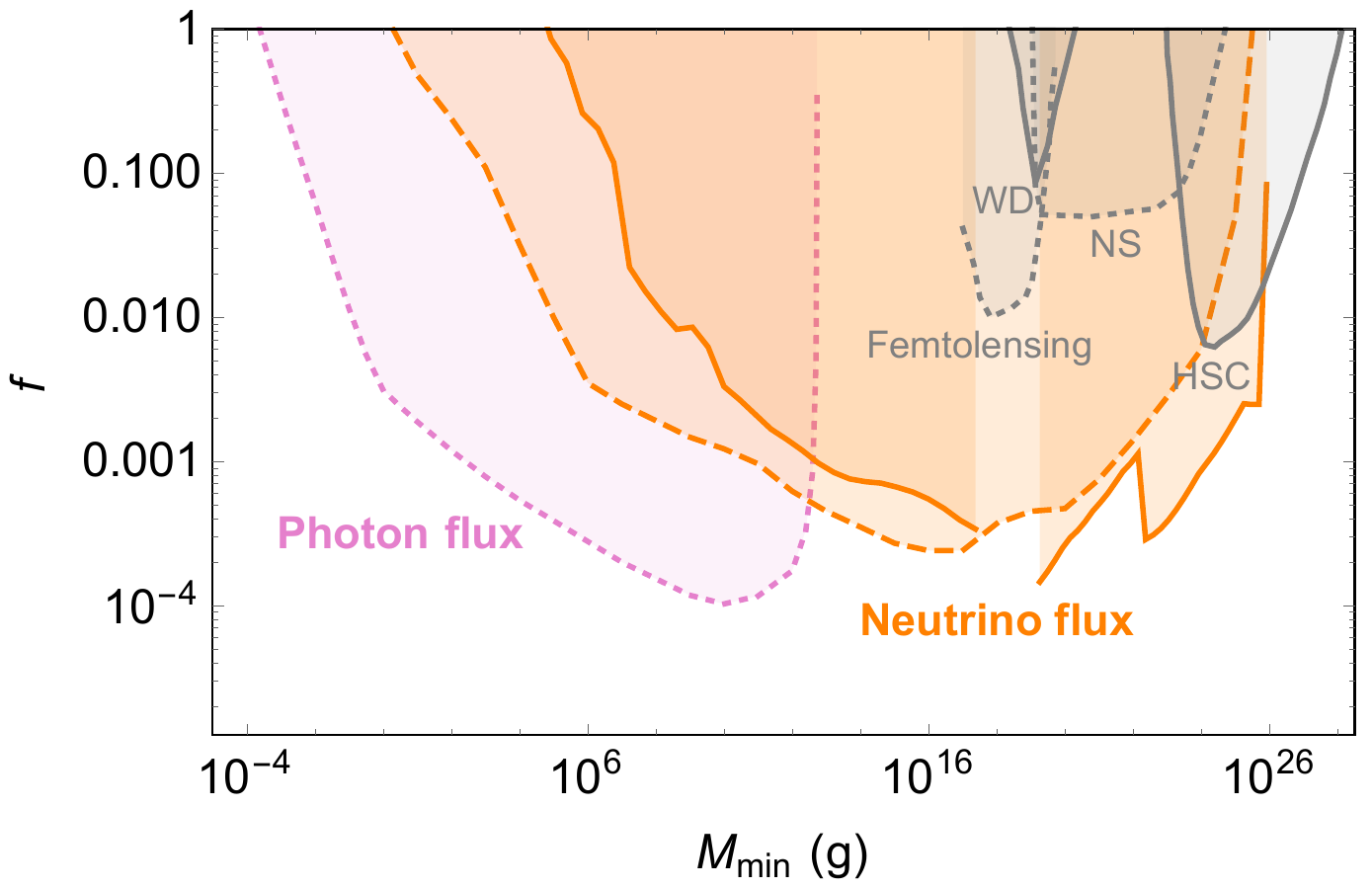}}
\caption{\label{fig:LUcons} 
Constraints on the present mass fraction of 2-2-hole remnants, $f\equiv \rho(t_0)/\rho_\textrm{DM}(t_0)$, as a function of $\Mmin$. The colored lines denote the constraints through the high-energy particle flux produced by binary mergers of 2-2-hole remnants. 
The orange dashed lines display the upper bound from neutrino observations by considering the parton shower of initial quarks, whereas the orange solid line includes only the on-shell production. The pink dotted line shows the constraints from the diffuse photon flux. The gray lines show a few upper bounds, as directly adapted from PBHs with the same mass, from femtolensing of gamma-ray bursts~\cite{Barnacka:2012bm}, the dynamical constraints from disruptions of white dwarfs (WD) and neutron stars (NS)~\cite{Graham:2015apa, Capela:2013yf}, and the microlensing observations, e.g. HSC~\cite{Niikura:2017zjd, Smyth:2019whb}. (Note that the validity of the gray dotted lines has recently been questioned~\cite{Katz:2018zrn, Conroy:2010bs, Ibata:2012eq}.) The plot is taken from~\cite{Aydemir:2020xfd}.}
\end{figure}

The extragalactic neutrino flux can be defined as~\cite{Carr:2009jm}
\begin{eqnarray}
\label{flux-EG}
\Phi_\nu^\textrm{EG}=\frac{c}{4\pi}\frac{n_{\nu }}{E_{\nu  }}
= \frac{c}{4\pi}\frac{n(t_0)}{E_{\nu}} \int^{t_0}_{t_\textrm{min}} E_{\nu}(t)  \frac{d N_\nu}{dE_\nu dt} e^{-S_\nu(E_\nu(t),z)}\;dt\,,
\end{eqnarray}
where $dt=-dz/[(1+z) H_0 \sqrt{(1+z)^3\Omega_M+(1+z)^4\Omega_R+\Omega_{\Lambda}}]$ and $E_{\nu }=E_\nu(t)/(1+z(t))$ is the redshifted energy. $n_{\nu }$ denotes the number density of neutrinos at present, where emissions extending back to $t_\textrm{min}$ (such that $E_\nu(t_\textrm{min})=\langle E_q \rangle$) are integrated over for a given $E_{\nu }$. $S_\nu(E, z)$ is the neutrino opacity factor~\cite{Gondolo:1991rn}. 
The results are summarized in Fig.~\ref{fig:LUcons} and will be discussed in section \ref{sec:results}.\\\\
\textit{Radiation from single remnants:}
The present observations can also probe the possible radiation from single remnants. For instance, the diffuse $\gamma$-ray background has been studied in the case of PBHs that haven't completed (or just completed) their evaporation by now, i.e. with mass  around $10^{14}\text{--}10^{15}$\,g~\cite{Page:1976wx,MacGibbon:1991vc,Carr:2009jm}. For 2-2-holes that become remnants in the very early universe, the photon background receives contribution from both the Milky way at present and the whole universe from early times. The problem is that the amount of radiation turn out to be too weak to be relevant to current data.

Since the temperature of remnants, as indicated in (\ref{eq:SMlimitTime}), is relatively weak, there is no need to take into account secondary production of photons. The galactic contribution to the flux can be computed as in (\ref{flux}) where the photon emission rate is given as
\begin{eqnarray}
\label{emissionrate}
 \frac{d N_{\gamma}}{dE_{\gamma} dt}\approx \frac{\pi^2}{60}\times A \frac{T_{\infty}^4}{\langle{E}_{\gamma}\rangle^2}\;.
\end{eqnarray}
where $A=4\pi r_{H}^2$. We take the emission energy as the average photon energy $\langle{E}_{\gamma}\rangle\approx 5.7 \,T_{\infty}$~\cite{Carr:2009jm}, which ranges from $0.1\,$keV to $10\,$MeV for $\Mmin=\Mp\;\text{--}\;10^{22}\,$g. The extragalactic contribution can be estimated through the corresponding version of (\ref{flux-EG}) where the Planckian distribution for the emission rate, as given in (\ref{emissionrate0}), is used in order to take into account the emission from the time of recombination and afterwards. It turns out that extragalactic and galactic contributions are in the same order of magnitude.

Including both galactic and extragalactic components, we find that a possible 2-2-hole contribution is too weak to be seen by the current existing data for isotropic photon flux~\cite{Strong:2004ry, Sreekumar:1997un}. The largest value of the anticipated flux that can be compared to the available data, which corresponds to the average photon energy $\braket{E_{\gamma}}\approx 0.7$ keV for $\Mmin\approx 10^{22}\textrm{ g}$, is six orders of magnitude smaller than the observed value. For smaller $\Mmin$, the contribution becomes even more insignificant, where it is fifteen orders of magnitude too small for $\Mmin\approx\Mp$ ($\braket{E_{\gamma}}\approx 15$ MeV).


\subsection{Early-universe constraints}
\label{sec:relic}


\textit{DM abundance:} If the primordial objects are subdominant in the energy budget before the instantaneous  evaporation, the resultant entropy injection is negligible, and $n(t)/s(t)$ remains constant till the present. This occurs if the number density at formation is smaller than a critical value,
\begin{eqnarray}
n(t_\textrm{init})\lesssim n_c(t_\textrm{init})=\frac{\rho_\textrm{tot}(t_\textrm{init})}{\Mini}\sqrt{\frac{t_\textrm{init}}{t_\textrm{eva}}}\,,
\end{eqnarray}
$t_\textrm{eva}\equiv t_\textrm{init}+\tau_L\approx \tau_L$ (or $\tau_\textrm{BH}$). In this case, from (\ref{eq:betaf0}) we have the following relation between the mass fraction for the remnant at present $f$ and the number density at formation with $n(t_0)/s(t_0)=n(t_\textrm{init})/s(t_\textrm{init})$,
\begin{eqnarray}\label{eq:betaf}
f=2.6\times 10^{28}\;\frac{\Mmin}{\Mp} \frac{n(t_\textrm{init})}{s(t_\textrm{init})}\,.
\end{eqnarray} 
Note that it receives a suppression factor of $\Mmin/\Mini$, in comparison to large PBHs. 

The situation is different if the primordial objects become dominant at some earlier time, where then there is a new era of matter domination that lasts until $t_\textrm{eva}$. In this case, we have a maximal value for the mass fraction that is obtained by replacing $n(t_\textrm{init})$ by $n_c(t_\textrm{init})$ in (\ref{eq:betaf}). In order for the 2-2-hole remnant to account for all of dark matter then requires this maximum value to be greater than unity, which imposes an upper bound on the initial mass as
\begin{eqnarray}\label{eq:Mrelic} 
\Mini\lesssim M_\textrm{DM}
\equiv 
6.8\times 10^5\,
\left(\frac{\Mmin}{\Mp}\right)^{4/5}\;
N^{-1/5}\,N_*^{1/5}\,g_*^{-1/10}\, \,\textrm{g}\,.
\end{eqnarray}
In comparison to PBHs, the bound is relaxed if $\Mmin$ is large, which corresponds to the weak coupling scenario.\\


\textit{BBN:} The effects of 2-2-holes on BBN can be investigated in connection with the corresponding analyses of PBHs evaporation that have been heavily studied in the literature~\cite{1978SvA....22..138V,10.1143/PTP.59.1012,1977SvAL....3..110Z,1978SvAL....4..185V,Kohri:1999ex,Carr:2009jm}. The underlying physics~\cite{Carr:2009jm} is naturally very similar in both cases. There are several ways that the emitted particles can affect BBN. First, in the beginning of BBN, at $t\sim 10^{-2}\text{--}10^{2}$\,s,  high energy mesons with long enough lifetime induce additional interconversion between protons and neutrons by scattering off the ambient nucleons, which alters the freeze-out value of $\textrm{n}/\textrm{p}$. Second, at $t\sim 10^2\text{--}10^{4}$\,s, high energy hadrons disassociate background nuclei before losing its energy through electromagnetic interaction. The target nuclei is predominantly ${^4}$He at this stage, due to the its high abundance. This leads to diminution of the ${^4}$He abundance and enhancement of the abundance of D, T, ${^3}$He, ${^6}$Li, and ${^7}$Li . Finally, high energy photons generated indirectly through scattering of the initial high energy quarks and gluons
cause further disassociation of ${^4}$He, enhancing the abundance of the lighter elements. This process, namely photodissociation, occurs at $t\sim 10^{4}\text{--}10^{12}$\,s.

The effects of radiating holes on the BBN processes are directly proportional to the emission rate $\Gamma_h(t)$  for hadronic particles, 
\begin{eqnarray}\label{eq:emissionrate}
\Gamma_h (t)= B_h \,n(t) \,\frac{1}{\langle E_h(t)\rangle} \frac{dM}{dt}\,,
\end{eqnarray}
where $B_h$ is the hadronic branching ratio and $\langle E_h (t)\rangle$ is roughly $T_\infty(t)$ up to an $O(1)$ factor.
Therefore, constraints on the mass fraction of primordial 2-2-holes can be inferred from the corresponding analysis in the PBH case~\cite{Carr:2009jm} by computing the ratio $\Gamma_{h,\textrm{BH}}/\Gamma_{h,22}$. During the relevant time-period, \textit{i.e.} $10^{-2}$ s $\lesssim t \lesssim 10^{12}$\,s, 2-2-holes with $M_\textrm{BBN} \lesssim \Mini \lesssim M_\textrm{rec}$ (as defined in (\ref{eq:Mi})) remain at the early stage of evaporation, where they radiate much like PBHs. By comparing the 2-2-hole in this stage to black hole with $\tau_L=\tau_{\textrm{BH}}$, we observe from (\ref{eq:LMlimitTime0}) that $\Gamma_i$ only differs by an overall constant and $\Gamma_{\textrm{BH}}/\Gamma_{22}$ is time independent. Therefore, BBN constraints for 2-2-holes can be found by a simple scaling of the corresponding constraints for PBHs as
\begin{eqnarray}
\label{betas}
\frac{n(t_\textrm{init})}{s(t_\textrm{init})}
\approx 0.4\;\hMini^{-3/2}\beta_\textrm{BH}\,\gamma^{1/2}g_*^{-1/4}\,  A^{1/3}\,,
\end{eqnarray}
where $A\equiv1.7\; N^{-1/4} \,\hMmin^{1/2}$ is the factor that appears in $T_\infty$ in (\ref{eq:LMlimit}).
Here, we convert the ratio of the number density to entropy density for black holes to the quantity $\beta_\textrm{BH}\,\gamma^{1/2}g_*^{-1/4}$, for which the latest constraints are given in \cite{Carr:2009jm}. On the other hand, 2-2-holes that are in the remnant stage in this time-period, i.e.  $\Mini\lesssim M_\textrm{BBN}$, produce much weaker radiation, which can be safely ignored. 

The results for the mass fraction at formation are displayed in Fig.~\ref{fig:EUcons}. The constraints are so strong that 2-2-holes cannot dominate the energy density before $t_\textrm{eva}$, i.e. $n(t_\textrm{init})<n_c(t_\textrm{init})$. Thus, the upper bound on the mass fraction at present can be found by using  (\ref{eq:betaf}), 
 %
 %
where BBN puts stronger constraints for  2-2 hole remnants in comparison to a PBH remnant with the mass $\Mmin$. 
For $\Mmin\lesssim10^{17}$ g, the maximum possible value of $f$, mentioned right above (\ref{eq:Mrelic}), is less than unity for $\Mini$ relevant for BBN constraints, and 2-2-hole cannot constitute all of dark matter independent of the BBN observations. It is for larger $\Mmin$ that there is a small region with this value being larger than unity, which is excluded directly by the BBN observations. Therefore, since we focus on the smaller $\Mmin$ range, the relevant bound is still $\Mini\lesssim M_\textrm{DM}$, as given in (\ref{eq:Mrelic}).


\textit{Entropy injection:}
For 2-2-holes that completed the early-time evaporation before BBN, i.e. $\Mini\lesssim M_\textrm{BBN}$, photons emitted are completely thermalized and only contribute to the density of background radiation. By using the observed baryon-to-photon ratio one can obtain an upper bound on the entropy injection from a possible 2-2-hole-dominant phase~\cite{1976Zel}, provided that the baryon asymmetry is generated purely in the early universe. The baryon number density is bounded from above by the radiation density right before the final evaporation, while the entropy density is bounded from below by the 2-2-hole entropy injection. Taking into account the upper bound on the  baryon-to-photon ratio $n_B/n_\gamma$, which should be larger than the observed value $6\times 10^{-10}$, we obtain an upper bound on the ratio of 2-2-hole number density to entropy density as
\begin{eqnarray}\label{eq:entropybound}
\frac{n(t_\textrm{init})}{s(t_\textrm{init})}\lesssim 1.2\times 10^7\,N^{-1/2}N_*^{1/4} \hMmin \, \hMini^{-5/2}\,.
\end{eqnarray}
Note that this constraint is weaker than the requirement of generating the observed relic abundance (for $\Mini\lesssim M_\textrm{DM}$), and is relevant only for $M_\textrm{DM}\lesssim \Mini \lesssim M_\textrm{BBN}$. 

On the other hand, the baryon asymmetry could also be generated by the evaporation of primordial objects, as been discussed for PBHs in \cite{Carr:1976zz, Zeldovich:1976vw, Toussaint:1978br,Turner:1979bt, Grillo:1980rt}. This requires the initial temperature to be above the electroweak scale, namely, $\Mini\lesssim 10^{12}\left(\Mmin/\textrm{g}\right)^{1/2}\,$g, which is comparable to $\Mini\lesssim M_\textrm{BBN}$ for $\Mmin$ of interest. Therefore, 2-2-holes that complete the early-time evaporation before the BBN era may account for the observed baryon asymmetry, in which case the entropy bound (\ref{eq:entropybound}) does not apply.

\textit{CMB constraints:} 
 The emission after BBN, but before the time of recombination, i.e. $M_\textrm{BBN}\lesssim \Mini\lesssim M_\textrm{rec}$,  can cause distortions in the CMB spectrum. Since this part of the parameter space is already strongly constrained by the BBN observations as previously discussed, these constraints are of less interest. The emission after recombination, on the other,  leads to the damping of small scale CMB anisotropies, providing a new constraint on the number density of 2-2-holes for $\Mini \gtrsim M_\textrm{rec}$. Given that the dominant contribution comes from the early stage of evaporation as before, the bounds would be similar to the PBH case, which were investigated in \cite{Zhang:2007zzh}. It turns out that the constraint on $n(t_\textrm{init})/s(t_\textrm{init})$ can be well approximated by a simple form,
\begin{eqnarray}
\label{CMB}
\frac{n(t_\textrm{init})}{s(t_\textrm{init})}\lesssim 3\times 10^{-80}\,\textrm{B}_e^{-1}N^{\,0.8}\,N_*^{-0.8}\hMmin^{-1.5}\,\hMini^{1.3}\,,
\end{eqnarray}
where $\textrm{B}_e$ is the branching ratio for electrons and positrons, which dominates the energy that goes into heating the matter~\cite{Carr:2009jm}. For an order of magnitude estimation, we use $\textrm{B}_e\approx 0.1$. As in the case for PBHs, the bound turns out to be quite strong, but it becomes weaker with increasing $\Mini$. More details can be found in the main paper~\cite{Aydemir:2020xfd}.

\subsection{Results}
\label{sec:results}

\begin{figure}[!h]
\captionsetup[subfigure]{labelformat=empty}
\hspace{-1.2cm}
\begin{tabular}{rrr}
\subfloat[\quad(a) $\Mmin=m_{\mathrm{Pl}}$]{\includegraphics[width=8cm]{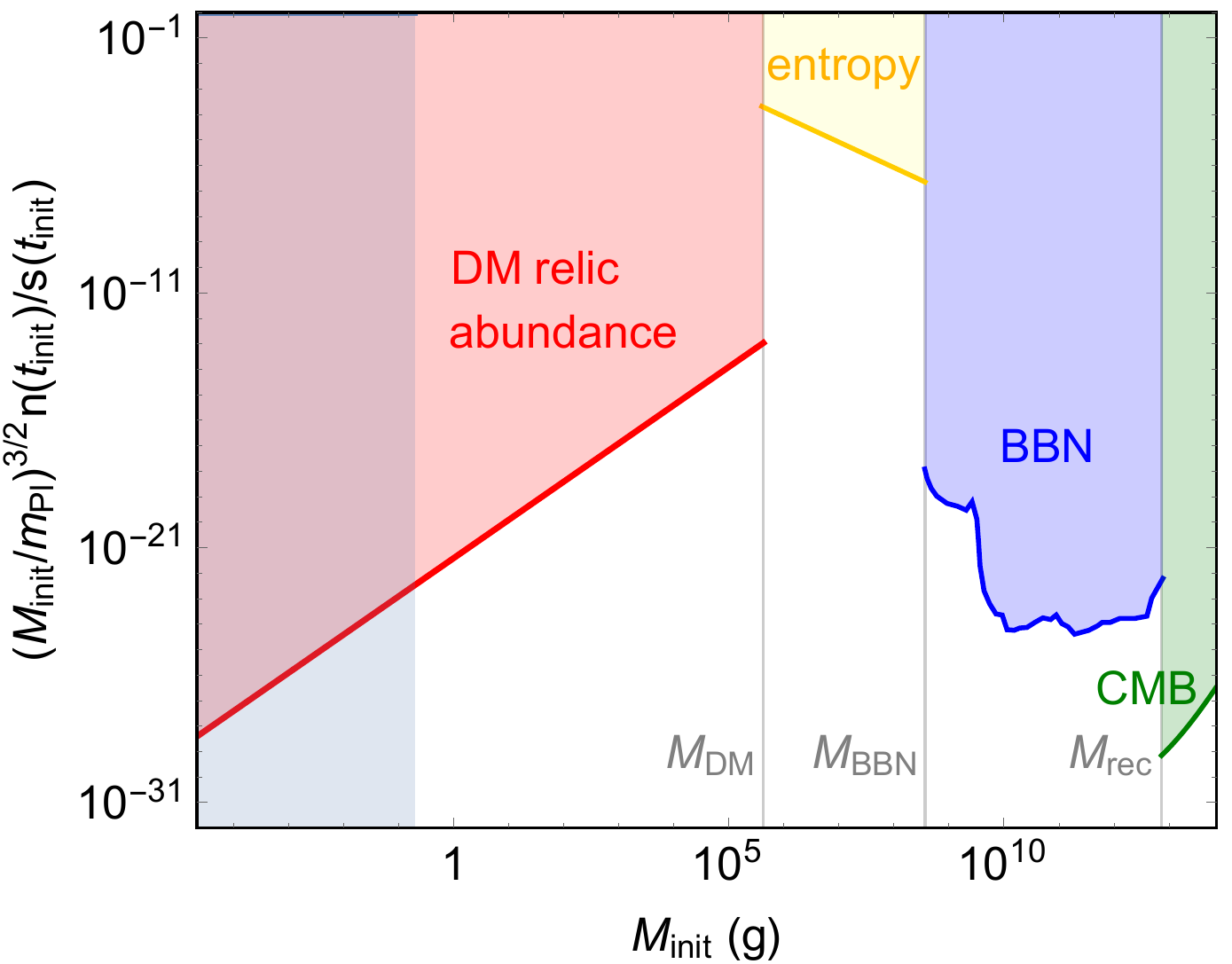}\label{fig:EU1}}&
\subfloat[\quad(b) $\Mmin=10^{20}$\,g]{\includegraphics[width=8cm]{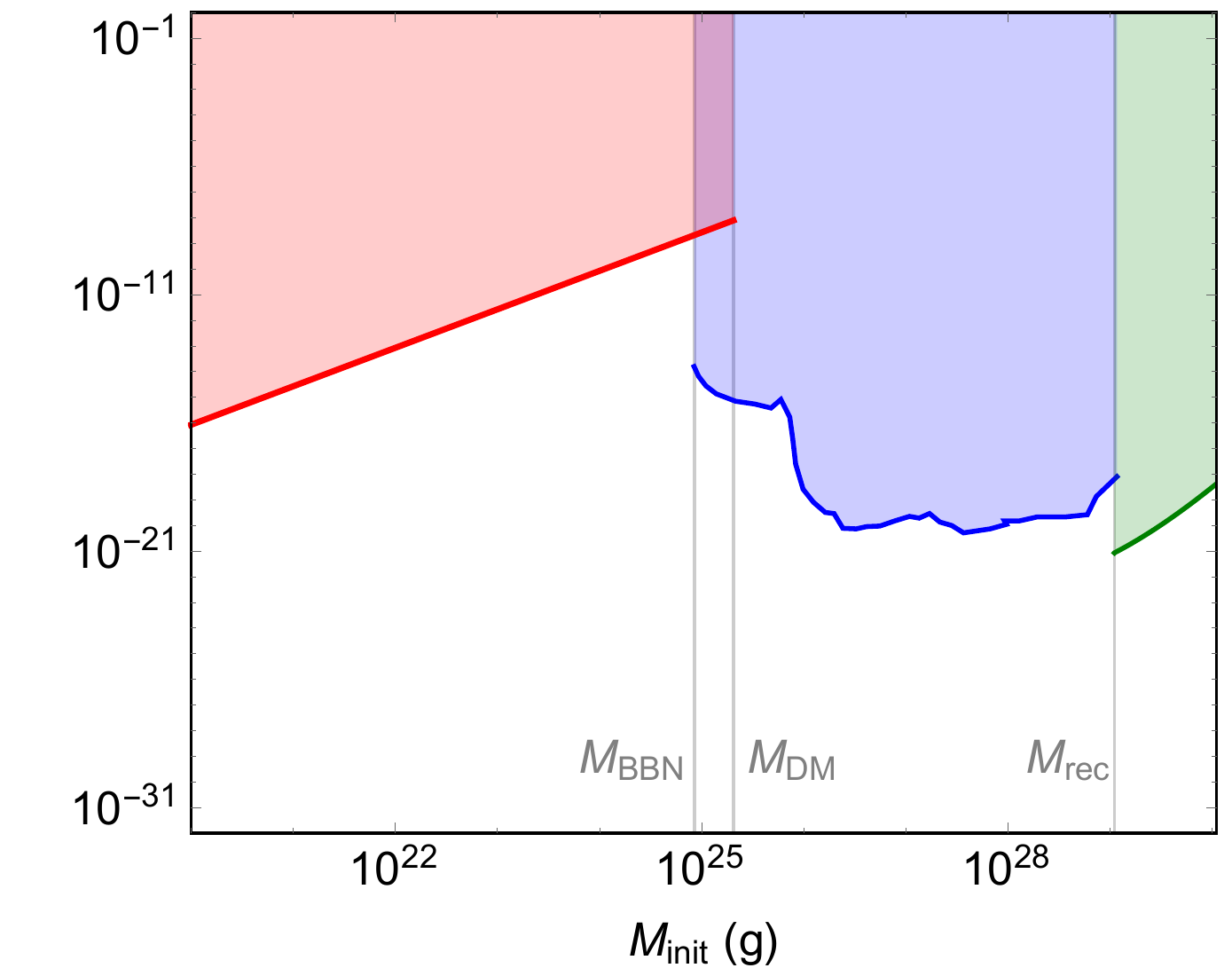}\label{fig:EU3}}
\end{tabular}
\caption{\label{fig:EUcons} 
The early-universe constraints as a function of $\Mini$ on the number density to entropy density ratio at formation $(\Mini/\Mp)^{3/2} \,n(t_\textrm{init})/s(t_\textrm{init})$,  which is related to the mass fraction at formation, $\beta$, as defined in (\ref{eq:beta}). The results are shown for two benchmark values of $\Mmin$.
We restrict to the small-mass range, where the primordial 2-2-holes have already become remnants today, i.e. $\Mmin \lesssim \Mini\lesssim M_\textrm{uni}$. 
The red region denotes the excluded parameter space by the observed relic abundance of dark matter. The yellow region shows the parameter space excluded due the photon-to-baryon ratio through entropy injection, which is valid only if the baryon asymmetry is not generated by the 2-2-hole evaporation. The blue and green regions display the exclusions from light element abundance formed in BBN and from CMB anisotropy, respectively.   
The critical masses $M_\textrm{BBN},\, M_\textrm{rec}$ in (\ref{eq:Mi}) and $M_\textrm{DM}$ in (\ref{eq:Mrelic}), specifing the mass ranges relevant to various observations, are shown by the gray vertical lines. 
The gray vertical band shows the range of $\Mini$ unavailable due to the minimal horizon mass in the radiation era, assuming the upper bound on the reheating temperature as $\sim10^{16}\,$GeV. Taken from~\cite{Aydemir:2020xfd}.}
\end{figure}

The results are summarized in Fig.~\ref{fig:LUcons} and Fig.~\ref{fig:EUcons}.  Fig.~\ref{fig:LUcons} presents the constraints on the present mass fraction ($f$) of 2-2-hole remnants in dark matter as a function of $\Mmin$. The constraints relevant to the mass range of interest result from high-energy particle fluxes produced by the merger product of the remnant binaries. On-shell production of neutrinos excludes remnants with $\Mmin\gtrsim 10^5\,$g from accounting for all of dark matter. The secondary production of neutrinos from parton showers of highly off-shell initial particles enables the exploration of the smaller range of $\Mmin$. By assuming the fragmentation functions in ordinary QCD, we find that the neutrino observations narrow the available parameter space down to $\Mmin\sim 1$\,g. Moreover, the photon flux data pushes the bound down to $\Mmin\lesssim  10\,\Mp$, although this may suffer from theoretical uncertainties from fragmentation function at small energy fraction and may be improved with a better understanding of parton showers for 2-2-hole evaporation.

Fig.~\ref{fig:LUcons} also shows the existing bounds from gravitational interaction, which become relevant for relatively large remnants, i.e. $\Mmin$ no smaller than $10^{17}\,$g. Since 2-2-holes gravitationally behave just like PBH, these bounds apply in both cases. In the case of PBH, the smaller mass range is considered untestable, since these objects are anticipated to have either evaporated away, or stopped evaporating and become remnants. The feature that a binary merger of two 2-2-hole remnants is not a remnant state but a hot 2-2-hole that evaporates like a black hole, hence radiating strongly due to its small mass, allows us to probe the small-mass range and hence opens a new window onto dark-matter parameter space.  

The early-universe constraints are displayed in Fig.~\ref{fig:EUcons}. The red region is excluded by the observed dark matter abundance (according to (\ref{eq:betaf}) with $f=1$), which constitutes the most relevant part of the constraints since this is the range ($\Mini\lesssim M_\textrm{DM}$) where remnants can account for all of dark matter, as discussed in Sec.~\ref{sec:relic}. 
The case for the Planck-mass remnants, corresponding to the strong coupling scenario for quadratic gravity, is given in Fig.~\ref{fig:EU1}. The allowed parameter space resembles closely the case of PBH relics with the Planck mass~\cite{Carr:2009jm, Dalianis:2019asr}. In this case, we have the condition $\Mini\lesssim M_\textrm{DM}\approx 4\times 10^5\,\textrm{g}$, meaning that the early stage of evaporation ends much before the BBN begins, since $\Mini\ll M_\textrm{BBN}$.  Larger 2-2-holes have too small remnant abundance
, but their number density can still be constrained by other requirements from the early universe; the photon-to-baryon ratio (\ref{eq:entropybound}), BBN (\ref{betas}), and CMB (\ref{CMB}) constraints. In the weak coupling scenario, where the remnants are heavier, the constraints in general differ from the black hole remnants. 
The parameter region constrained by entropy injection shrinks, and for $\Mmin\gtrsim 10^{17}\,$g we have $M_\textrm{DM}\gtrsim M_\textrm{BBN}$, thus the red and blue regions overlap. Therefore, in this case the parameter space for $\Mini$ for 2-2-hole remnants making up all of dark matter starts to be excluded by BBN observations. An example for this case is shown in Fig.~\ref{fig:EU3}. On the other hand, for $\Mmin\lesssim10^{17}\,$g in the weak coupling case, $\Mini\lesssim M_\textrm{DM}$ is still the determining constraint. For the $\Mmin$ bounds obtained above, this yields $\Mini\lesssim 10^{13}$\,g for $\Mmin\lesssim10^5$\,g and $\Mini\lesssim 10^{6}\,$g for $\Mmin\lesssim 10\,\Mp$.   



\section{Conclusion}
\label{sec:final}

Remnants from primordial thermal 2-2-holes constitute a well-motivated and promising candidate for dark matter. The fact that these remnants naturally arise in the theory puts them in a more compelling position over PBH remnants. Moreover, the 2-2-hole, being a probable endpoint of gravitational collapse instead of the black hole, offers a resolution to the information loss conundrum due to the absence of a horizon. 


In this work~\cite{Aydemir:2020xfd}, we have considered these remnants as dark matter and explored the corresponding astrophysical and cosmological implications. By taking into account the observational constraints, we have shown that there exists a viable region of parameter space accommodating these objects as all dark matter. 

 
 The 2-2-hole formation mass is constrained by the early-universe cosmology and bounded from above for a given remnant mass $\Mmin$, mainly by the requirement of generating the observed dark matter abundance; in consequence, the early stage of evaporation is required to end before BBN begins. The parameter space for $\Mmin$, on the other hand, can be probed by the present epoch observations, mainly of high-energy astrophysical particles, in addition to the conventional PBH searches through gravitational interactions. A binary merger of two remnants, which is not a remnant state but a hot 2-2-hole, generates a strong flux of high energy particles almost instantly before settling back down to a cold remnant. The neutrino bounds, being less susceptible against theoretical uncertainties in the fragmentation function from parton showers, yield $\Mmin\lesssim 10^5\,$g as a conservative estimate. The photon observations provide stronger bounds, strengthening the constraint further to be $\Mmin\lesssim 10\,\Mp$. As a result, our findings point towards a strong-coupling regime for the theory of quantum gravity, and hence a single fundamental scale.

\section*{Acknowledgements} 
\vspace{-0.2cm}
U.A. thanks the organisers of Corfu Summer Institute 2019 "School and Workshops on Elementary Particle Physics and Gravity" (CORFU2019) for the invitation. U.A. also thanks Bob Holdom and Jing Ren for comments on the manuscript. Work of U.A. is supported in part by the Chinese Academy of Sciences President's International Fellowship Initiative (PIFI) under Grant No. 2020PM0019, and the Institute of High Energy Physics, Chinese Academy of Sciences, under Contract No.~Y9291120K2. 

\bibliography{References_22_holes}
\bibliographystyle{JHEP}

\end{document}